\documentclass[letter,traditabstract,letterpaper]{aa}
\usepackage{graphicx,txfonts,natbib}
\begin{document}

\title{{\em Herschel}/PACS Imaging of Protostars in the\\HH 1--2 Outflow Complex\thanks{{\it Herschel} is an ESA space observatory with science instruments provided by European-led Principal Investigator consortia and with important participation from NASA. This work includes data acquired with the Atacama Pathfinder Experiment (APEX; E-082.F-9807, E-284.C-5015). APEX is a collaboration between the Max-Planck-Institut f\"ur Radioastronomie, the European Southern Observatory, and the Onsala Space Observatory.}}

\author{W. J. Fischer \inst{1}
   \and S. T. Megeath \inst{1}
   \and Babar Ali \inst{2}
   \and J. J. Tobin \inst{3}
   \and M. Osorio \inst{4}
   \and L. E. Allen \inst{5}
   \and E. Kryukova \inst{1}
   \and T. Stanke \inst{6}
   \and A. M. Stutz \inst{7,8}
   \and E. Bergin \inst{3}
   \and N. Calvet \inst{3}
   \and J. Di~Francesco \inst{9,10} 
   \and E. Furlan \inst{11}
   \and L. Hartmann \inst{3}
   \and T. Henning \inst{7}
   \and O. Krause \inst{7}
   \and P. Manoj \inst{12}
   \and S. Maret \inst{13}
   \and J. Muzerolle \inst{14}
   \and P. Myers \inst{15}
   \and D. Neufeld \inst{16}
   \and K. Pontoppidan \inst{17}
   \and C. A. Poteet \inst{1}
   \and D. M. Watson \inst{12}
   \and T. Wilson \inst{6}
   }

\institute{Department of Physics and Astronomy, University of Toledo, 2801 West Bancroft Street, Toledo, OH 43606, USA\\
              \email{wfische@utnet.utoledo.edu}
   \and NHSC/IPAC/Caltech, 770 South Wilson Avenue, Pasadena, CA 91125, USA
   \and Department of Astronomy, University of Michigan, 830 Dennison Building, 500 Church Street, Ann Arbor, MI 48109, USA
   \and Instituto de Astrofisica de Andalucia, CSIC, Camino Bajo de Huetor 50, E-18008, Granada, Spain
   \and National Optical Astronomy Observatory, 950 North Cherry Avenue, Tucson, AZ 85719, USA 
   \and ESO, Karl-Schwarzschild-Strasse 2, 85748 Garching bei M\"{u}nchen, Germany
   \and Max-Planck-Institut f\"{u}r Astronomie, K\"{o}nigstuhl 17, D-69117 Heidelberg, Germany
   \and Department of Astronomy and Steward Observatory, University of Arizona, 933 North Cherry Avenue, Tucson, AZ 85721, USA 
   \and Department of Physics and Astronomy, University of Victoria, P.O. Box 355, STN CSC, Victoria BC, V8W 3P6, Canada
   \and National Research Council Canada, Herzberg Institute of Astrophysics, 5071 West Saanich Road, Victoria BC, V9E 2E7, Canada
   \and {\it Spitzer} Fellow; Jet Propulsion Laboratory, Caltech, Mail Stop 264Ð767, 4800 Oak Grove Drive, Pasadena, CA 91109, USA
   \and Department of Physics and Astronomy, University of Rochester, Rochester, NY 14627, USA 
   \and Laboratoire d'Astrophysique de Grenoble, Universit\'e Joseph Fourier, CNRS, UMR 571, BP 53, F-38041 Grenoble, France
   \and Space Telescope Science Institute, 3700 San Martin Drive, Baltimore, MD 21218, USA
   \and Harvard-Smithsonian Center for Astrophysics, 60 Garden Street, Cambridge, MA 02138, USA
   \and Johns Hopkins University, 3400 North Charles Street, Baltimore, MD 21218, USA 
   \and Division of Geological and Planetary Sciences 150-21, California Institute of Technology, Pasadena, CA 91125, USA
   }

   \date{}

   \abstract{We present 70 and 160 $\mu$m {\it Herschel} science demonstration images of a field in the Orion A molecular cloud that contains the prototypical Herbig-Haro objects HH 1 and 2, obtained with the Photodetector Array Camera and Spectrometer (PACS).  These observations demonstrate {\it Herschel}'s unprecedented ability to study the rich population of protostars in the Orion molecular clouds at the wavelengths where they emit most of their luminosity.  The four protostars previously identified by {\it Spitzer} 3.6 -- 40 $\mu$m imaging and spectroscopy are detected in the 70 $\mu$m band, and three are clearly detected at 160 $\mu$m.  We measure photometry of the protostars in the PACS bands and assemble their spectral energy distributions (SEDs) from 1 to 870 $\mu$m with these data, {\it Spitzer} spectra and photometry, 2MASS data, and APEX sub-mm data.  The SEDs are fit to models generated with radiative transfer codes.  From these fits we can constrain the fundamental properties of the protostars.  We find luminosities in the range 12 -- 84 $L_{\sun}$ and envelope densities spanning over two orders of magnitude.  This implies that the four protostars have a wide range of envelope infall rates and evolutionary states: two have dense, infalling envelopes, while the other two have only residual envelopes.  We also show the highly irregular and filamentary structure of the cold dust and gas surrounding the protostars as traced at 160 $\mu$m.}

   \keywords{Stars: formation --- Stars: protostars --- circumstellar matter --- Infrared: ISM --- Infrared: stars}
   
   \maketitle

\section{Introduction}

The Orion molecular clouds are the most active region of star formation within 500 pc of the Sun, where the {\it Spitzer Space Telescope} identified over 400 likely protostars in the Orion A and B clouds (Megeath et al., in prep). The region is home to both clustered and distributed star formation and hosts both high- and low-mass protostars.  The {\it Herschel Space Observatory}'s capabilities in the far infrared are crucial for sampling the expected peak of the spectral energy distributions (SEDs) of protostars, which are dominated by thermal emission from a cold ($\sim$10~K) envelope.  Measuring the peak of the SED allows firm estimates of the bolometric luminosities and envelope densities of the protostellar systems.

With the Photodetector Array Camera and Spectrometer (PACS; \citealt{pog10}) aboard {\it Herschel} \citep{pil10}, we have obtained 70 and 160 $\mu$m images of a field in the Lynds 1641 region.  The field contains the intermediate-mass Herbig B9e star V380 Ori \citep{hil92}, 28 infrared excess sources identified by observations with {\it Spitzer} (Megeath et al., in prep.), and a variety of outflow phenomena including the well-known HH 1 and 2 \citep[e.g.,][]{bal02} and $\ge8$ protostellar outflows \citep[e.g.,][]{sta02}.  This is the science demonstration field for the {\it Herschel} Orion Protostar Survey (HOPS), a 200-hour open-time key program that will obtain PACS imaging of 133  fields, $5'$ to $8'$ in diameter, containing 278 protostars and PACS spectroscopy of a subset of 37 protostars.

Here, we present photometry of the four protostars in the {\it Herschel} field at 70 and 160 $\mu$m and combine these data with {\it Spitzer} and ground-based data.  With the radiative transfer code of \citet{whi03}, we generate model SEDs and find that the four protostars exhibit a large range of luminosities ($12<L/L_{\sun}<84$) and envelope densities spanning over two orders of magnitude.  This implies that two protostars have dense, infalling envelopes, while the other two have only residual envelopes.

\section{Observations and Data Reduction}

An $8'$ square field with central coordinates $\alpha=5^h36^m22^s.05$, $\delta=-6^\circ45'41''.23$ (J2000) was observed on 2009 October 9 (observing day 148; observation IDs 1342185551 and 1342185552) in the 70 $\mu$m (``blue") and 160 $\mu$m (``red'') bands available with PACS, which have angular resolutions of 5.2\arcsec\ and 12\arcsec, respectively.  We observed our target field with homogeneous coverage using two orthogonal scanning directions and a scan speed of $20''$/s.  Each scan was repeated 5 times for a total observation time of 1468~s per scan direction.  The effective sampling rate of the detectors is 10 Hz.  The data were processed from raw telemetry to final images with the {\it Herschel} Common Software System (HCSS) version 3.0 build 919, using version 4 of the flux calibration files.\footnote{HCSS is a joint development by the {\it Herschel} Science Ground Segment Consortium, consisting of ESA, the NASA {\it Herschel} Science Center, and the HIFI, PACS, and SPIRE consortia.}  We followed the standard processing steps for PACS data described by \citet{pog10} with these exceptions: We identified cosmic rays for each spatial sky pixel as those values which were larger than 10 standard deviations from the mean signal.  Several extraneous calibration measurements were interspersed with the HOPS target observations.  These were masked and removed from the data cube with an additional 430 readouts following each of these calibration measurements to mask signal drifts induced by the calibration source.

After the initial processing, the two orthogonal scan observations were combined for the final mapmaking step.  Two different mapping approaches are used for this purpose: Method~1 is used exclusively for point source photometry, while Method~2 is used to display images.

Method~1: Mapping with local sky subtraction: First, we remove the signal drifts (whether correlated or due to the $1/f$ noise) by subtracting a local ``sky'' value from each readout from each bolometer pixel.  The local sky is estimated as the median value within a window of size $\pm20$ readouts.  The final mosaic is then created by spatially averaging all overlapping bolometer pixels using the HCSS routine ``photProject.''Ê To protect the integrity of the point source PSF, all readouts within 20\arcsec\ of a point source are ignored during the sky median calculation.  This processing preserves all point and compact sources in the image and provides the proper photometry comparison between HOPS target objects and the flux calibration standards, which use the same reduction scheme.  However, this processing removes all emission at spatial scales larger than the median window size.

Method~2: Mapping without local sky subtraction: Method~1 is necessary only for accurate photometry of point sources. ÊWe also create maps by removing only the pixel-to-pixel electronic offsets in PACS images, using the median value of the entire time stream of a single pixel to estimate its offset signal value.  Unlike Method~1, this approach does not remove the (spatially) extended emission.  However, it also does not mitigate the $1/f$ drifts, which add so-called ``striping'' or ``banding'' in the final maps.  As for Method~1, we use the ``photProject'' HCSS routine to spatially coadd individual array readouts for mapmaking.

A $K_s$ image of the field was acquired with NEWFIRM, the NOAO Extremely Wide Field Infrared Imager, on the KPNO 4 m telescope, and the data were reduced with the NOAO NEWFIRM Pipeline \citep{swa09}.  The on-source time was 11 minutes over most of the field of view.  Images at 350 and 870 $\mu$m were acquired at APEX with SABOCA and LABOCA, respectively.  The observing and data reduction procedures for the APEX images are described in \citet{sta10}.

\begin{figure} 
\resizebox{\hsize}{!}{\includegraphics{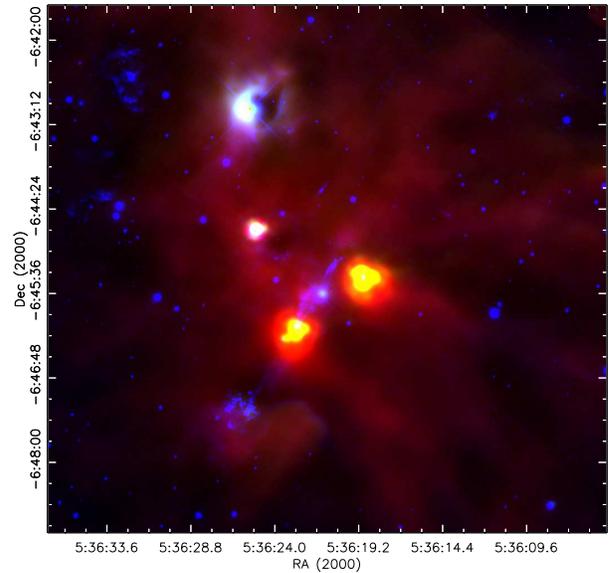}} 
\caption{\small Three-color composite image of the HH 1--2 region.  Blue is NEWFIRM $K_s$, green is PACS 70 $\mu$m, and red is PACS 160 $\mu$m.} 
\label{f.3color} 
\end{figure} 

\section{Results}

\subsection{Imaging and Photometry}

Figure~\ref{f.3color} shows a composite of the final map created using Method~2 for the 70 and 160 $\mu$m PACS channels.  Figures~\ref{f.blue} and \ref{f.red}, available in electronic form only, show the separate 70 and 160 $\mu$m images and are annotated with source names.

\onlfig{2}{
\begin{figure*}
\resizebox{\hsize}{!}{\includegraphics{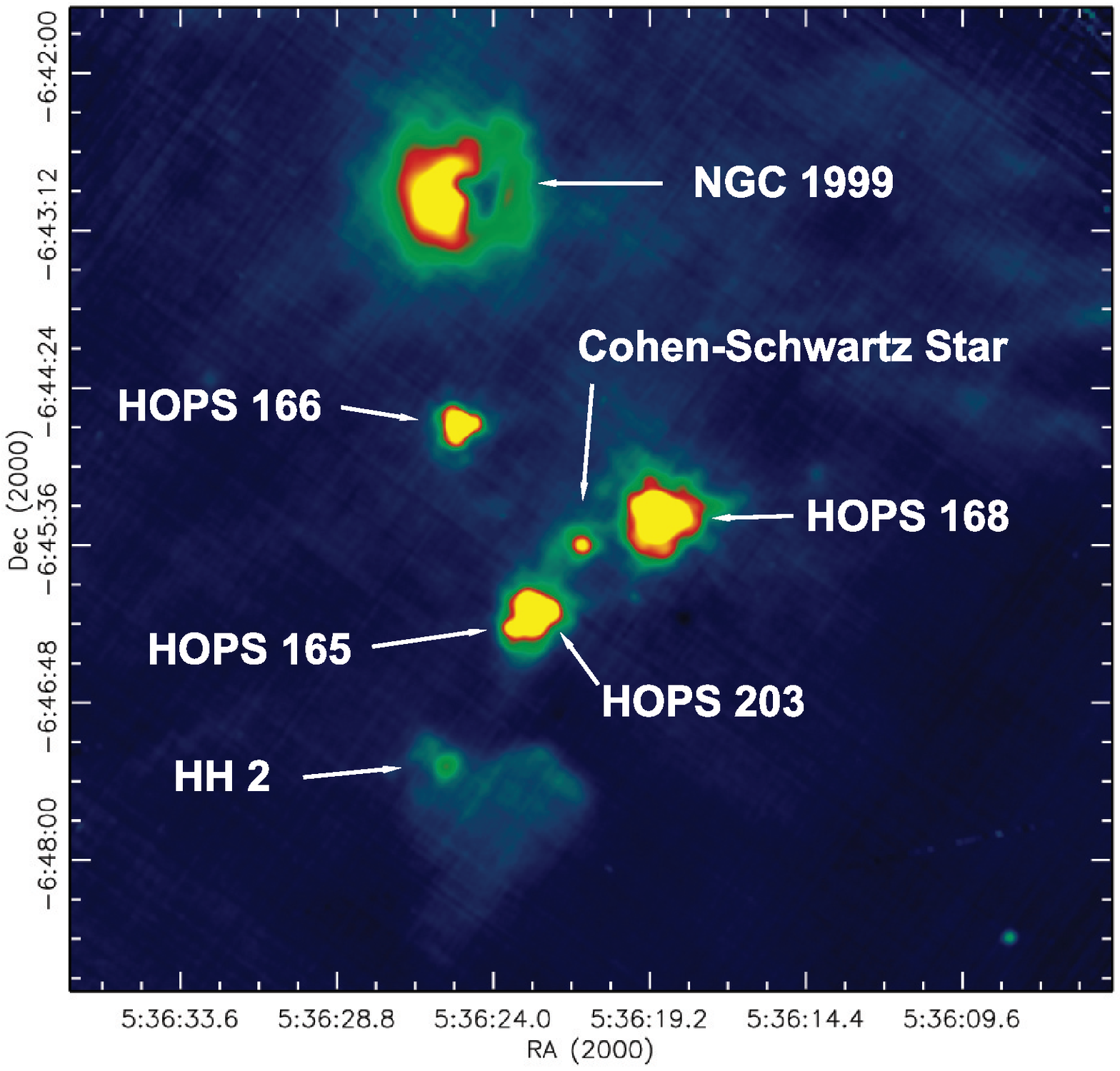}}
\caption{\small PACS 70 $\mu$m image of the HH 1--2 region.}
\label{f.blue}
\end{figure*}
}

\onlfig{3}{
\begin{figure*}
\resizebox{\hsize}{!}{\includegraphics{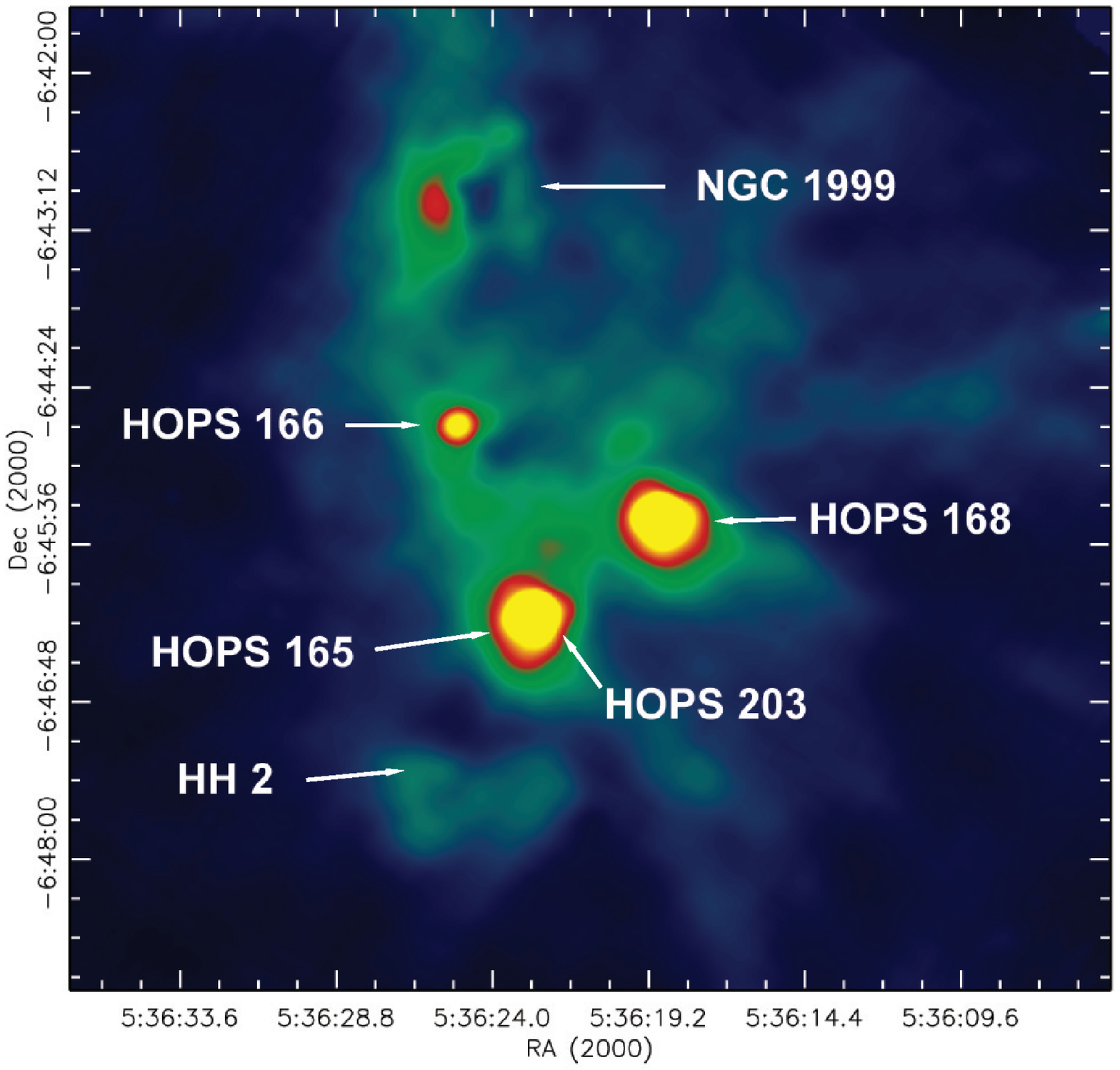}}
\caption{\small PACS 160 $\mu$m image of the HH 1--2 region.}
\label{f.red}
\end{figure*}
}

The bright blue source in the north of the field is the reflection nebula NGC 1999; the dark tri-lobed feature seen toward this nebula is discussed in \citet{sta10}.  In the center of the image is a triangular arrangement of protostars.  Here we use their designations for the HOPS program: 165, 166, 168, and 203.  HOPS 166 (HH 147 MMS; Chini et al. 2001) is the relatively isolated source at the northeastern corner of the triangle, HOPS 168 (HH 1--2 MMS 2) is at the western corner, and HOPS 165 and 203 are the pair of overlapping sources (separated by $13''$) at the southern corner.  HOPS 203 (HH 1--2 MMS 1), the brighter of the pair in the PACS bands, is the source of the HH 1--2 outflow and contains the radio sources VLA 1 and 2 \citep{rod90}.  \citeauthor{chi01} report an additional source HH 1--2 MMS 3, 22\arcsec\ southwest of HOPS 168, that corresponds to extended emission at 160 $\mu$m with no apparent point source at 70 $\mu$m.  Falling nearly along the line between HOPS 168 and 203 is the C-S star, a classical T Tauri star \citep{coh79}.  To the southeast of HOPS 203 at $\alpha=5^h36^m25^s.3$, $\delta=-6^\circ47'18''$ is a knot of emission presumably shock heated by the HH 2 outflow.  At 160 $\mu$m, only HOPS 166, 168, and 203 appear, while HOPS 165 is not detected.  The 160 $\mu$m band also traces cold dust in the surrounding cloud material, showing an irregular, filamentary structure.

PACS photometry of the four protostars appears in columns 9 and 10 of Table~\ref{t.photo}. We obtained simple aperture photometry for the relatively isolated protostars: HOPS 166 and 168 in the blue and red bands and HOPS 203 in the red band.  In these cases, we used a 16\arcsec\ aperture with subtraction of the median signal in a background annulus extending from 18\arcsec\ to 22\arcsec.  The results were corrected according to measurements of the encircled energy fraction provided by the PACS consortium (priv.\ comm.).

For the HOPS 165/203 pair at 70 $\mu$m, point-spread function (PSF) fitting was required to separate the fluxes of the two protostars.  We fit the fainter HOPS 165 with a PSF constructed from observations of Vesta (PACS consortium, priv.\ comm.).  Aperture photometry for HOPS 165 was performed on the best-fit PSF, and aperture photometry for HOPS 203 was performed on the data after subtraction of the HOPS 165 model.  At 160 $\mu$m, we report an upper limit for HOPS 165; this is the largest flux density for which a model PSF can be added at the source position before it appears as an asymmetry in the HOPS 203 image.

According to \citet{pog10}, the calibration accuracy for PACS is within 10\% in the blue band and better than 20\% in the red.  The formal uncertainties associated with each source (i.e., the RMS of the signal in the sky annulus) are much less, $\le1$\%, except for the case of HOPS 165, where fitting a point-spread function to a faint source yields a 10\% uncertainty.

The PACS photometry data are supplemented by {\it Spitzer} IRAC and MIPS photometry (Megeath et al., in prep.), {\it Spitzer} IRS spectroscopy, and APEX SABOCA and LABOCA sub-mm photometry \citep{sta10}.  For HOPS 166, near-infrared J/H/K photometry was available from the Two Micron All Sky Survey.\footnote{The Two Micron All Sky Survey (2MASS) is a joint project of the University of Massachusetts and the Infrared Processing and Analysis Center/California Institute of Technology, funded by NASA and the National Science Foundation.}  The {\it Spitzer} positions and 3.6 -- 870 $\mu$m photometry for the HOPS protostars appear in Table~\ref{t.photo}.  Systematic uncertainties are given in a note to the table.

\begin{table*}
\caption{Protostar Photometry}
\label{t.photo}
\centering
\begin{tabular}{l cc cccc c cc cc}
\hline\hline
HOPS & RA (J2000) & Dec. (J2000) & [3.6] & [4.5] & [5.8] & [8.0] & [24] & [70] & [160] & [350] & [870] \\
Source & (h m s) & ($^\circ$ \arcmin\ \arcsec) & (Jy) & (Jy) & (Jy) & (Jy) & (Jy) & (Jy) & (Jy) & (Jy) & (Jy) \\  
\hline
165\dotfill & 5 36 23.54 & $-$6 46 14.6 & 0.014 & 0.052 & 0.11 & 0.13 & 0.64 & 1.1 & $<$0.3 & $<$3 & ... \\
166\dotfill & 5 36 25.13 & $-$6 44 41.8 & 0.66 & 0.83 & 0.97 & 1.17 & 4.65 & 10.9 & 11.1 & 4.6 & 0.33 \\
168\dotfill & 5 36 18.93 & $-$6 45 22.7 & 0.0077 & 0.030 & 0.041 & 0.038 & 3.66 & 87.3 & 87.7 & 24 & 0.94 \\
203\dotfill & 5 36 22.84 & $-$6 46 06.2 & ... & ... & ... & 0.0091 & 0.54 & 26.6 & 75.7 & 28 & 1.2 \\
\hline
\end{tabular}
\tablefoot{Aperture radii are 3\arcsec\ from 1.2 to 2.2 $\mu$m (not shown), 2\arcsec\ from 3.6 to 8.0 $\mu$m, 5\arcsec\ at 24 $\mu$m, and 16\arcsec\ at 70 $\mu$m and longward.  Systematic uncertainties are 3\% from 1.2 to 2.2 $\mu$m, 5\% from 3.6 to 24 $\mu$m, 10\% at 70 $\mu$m, 20\% at 160 $\mu$m, 40\% at 350 $\mu$m, and 20\% at 870 $\mu$m.}
\end{table*}

\subsection{SED Modeling}

We use a Monte Carlo radiative transfer code \citep{whi03} to calculate model SEDs for the four protostars.  The code features a central star and flared disk, which emit photons that can then be scattered or absorbed and re-emitted by dust in either the disk or an envelope.  The envelope density is defined by the rotating collapse solution of \citet{ter84}, plus a bipolar, evacuated cavity.

We use the same dust model as \citet{tob08}, which contains larger dust grains than a standard ISM dust model. The grain size distribution is defined by a power law $n\left(a\right)\propto a^{-3.5}$, with $0.005~\mu{\rm m} \le a \le 1~\mu{\rm m}$. We use dust grains composed of graphite $\zeta_{\rm graph}=0.0025$, silicates $\zeta_{\rm sil}=0.004$, and water ice $\zeta_{\rm ice}=0.0005$; abundances ($\zeta$) are relative to gas and imply a gas to dust ratio of 133. Our sub-mm opacities exceed those of the well-known Milky Way Case B ($R_V=5.5$) mixture of \citet{wei01} by a factor that reaches a maximum of 5 at 600 $\mu$m.

The model parameters are set to typical values for low-mass protostars; we fit the SEDs by varying seven of them: the system luminosity $L$, the reference envelope density $\rho_1$ \citep{ken93}, the outer radius of the envelope $R_{\rm env}$, the opening angle of the envelope cavity $\theta_{\rm cav}$, the mass of the disk $M_{\rm disk}$, the inclination angle $i$, and the foreground extinction $A_V$.  (Foreground extinction is applied with the laws of \citealt{mcc09}, suitable for star-forming regions.)  In fitting the sources, we emphasize the mid to far IR over the near IR, since the near IR is highly dependent on the scattering properties of the dust, the geometry of the inner disk, and the geometry of the outflow cavity.  Thus, when fitting a source we first adjust the luminosity and density to get the best fit to the mid to far IR, then we find the best combination of cavity opening angle, inclination, and (if necessary) foreground reddening to fit the 10 $\mu$m absorption feature and the near-IR emission.  In general, the fits are insensitive to $R_{\rm env}$ and $M_{\rm disk}$.  However, for HOPS 165, it was necessary to adjust these two parameters.  The best-fit parameters were determined by visual comparison of the models to the observed SEDs and are listed in Table~\ref{t.model}.

\begin{figure}
\resizebox{\hsize}{!}{\includegraphics{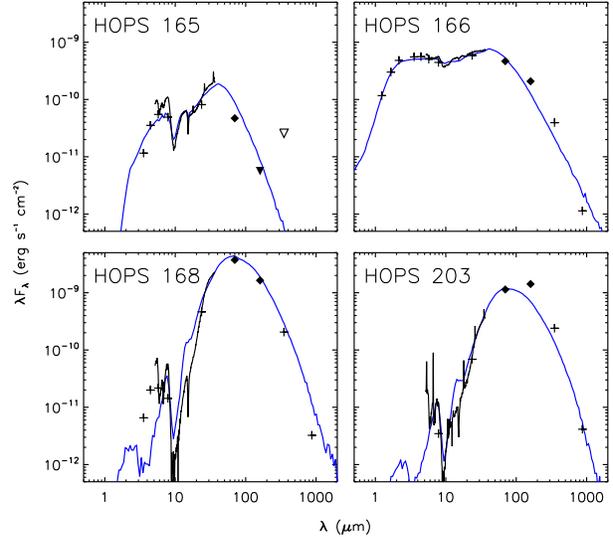}} 
\caption{\small Observed SEDs (black) with best-fit models (blue).  Plus signs indicate 2MASS, {\it Spitzer}, and APEX  photometry; the open triangle is an upper limit.  Black curves show {\it Spitzer} spectra.  Filled symbols indicate {\it Herschel} photometry: diamonds represent detections, while the filled triangle is an upper limit.\label{f.seds}}
\end{figure}

The models are compared to the photometry and spectra in Figure~\ref{f.seds}.  For these models, $4\times10^7$ photons were run through the Monte Carlo code.  The code generates output for apertures ranging from 1\arcsec\ to 16\arcsec\ in one-arcsecond steps, and the choice of aperture for the plotted SED varies with wavelength, as given in the note to Table~\ref{t.photo}. An interpolation scheme bridges the gaps between disparate apertures.  We assume a distance of 420 pc \citep{men07}.

\begin{table}
\caption{Adopted Model Parameters}
\setlength{\tabcolsep}{0.075in}
\label{t.model}
\centering
\begin{tabular}{l ccccccc}
\hline\hline
 HOPS & $L$ & $\rho_1^{\mathrm{a}}$ & $R_{\rm env}$ & $\theta_{\rm cav}$ & $M_{\rm disk}$ & $i$ & $A_V$ \\
 Source & ($L_{\sun}$) & (g cm$^{-3}$) & (AU) & ($^\circ$) & ($M_{\sun}$) & ($^\circ$) & \\
\hline
165\dotfill & 12 & $7.5\times10^{-16}$ & $10^3$ & 30 & $1\times10^{-4}$ & 20 & 35 \\
166\dotfill & 23 & $1.5\times10^{-15}$ & $10^4$ & 25 & $5\times10^{-2}$ & 40 & 4 \\
168\dotfill & 84 & $3.0\times10^{-13}$ & $10^4$ & 40 & $5\times10^{-2}$ & 75 & 0 \\
203\dotfill & 23 & $2.6\times10^{-13}$ & $10^4$ & 40 & $5\times10^{-2}$ & 75 & 0 \\
\hline
\end{tabular}
\begin{list}{}{}
\item[$^{\mathrm{a}}$]The envelope density at 1 AU in the limit of no rotation.
\end{list}
\end{table}

\section{Discussion}

The {\it Herschel} observations have detected four protostars in the HH 1--2 region. ÊAll were previously identified with {\it Spitzer} (Megeath et al., in prep), but {\it Herschel} has cleanly separated the sources for the first time in the far IR, allowing accurate photometry.  From the luminosities and densities in Table~\ref{t.model}, we estimate the infall rate from the envelope onto the central star-disk system, the luminosity due to accretion onto the star, and the evolutionary state for each protostar.  \citet{ken93} show that for a protostellar envelope, the infall rate is $\dot{M}_{\rm env}=1.9\times10^{-7}~(\rho_1/10^{-15}~{\rm g~cm}^{-3})~(M_*/M_{\sun})^{1/2}~M_{\sun}~{\rm yr}^{-1}$.  The accretion luminosity can then be written as $L_{\rm acc}=GM_*\dot{M}_{\rm disk}/R_*=5.9~(\rho_1/10^{-15}~{\rm g~cm}^{-3})~(M_*/M_{\sun})^{3/2}~(R_*/R_{\sun})^{-1}~(\dot{M}_{\rm disk}/\dot{M}_{\rm env})~L_{\sun}$, where $\dot{M}_{\rm disk}$ is the accretion rate from the disk onto the star.  To Êestimate $\dot{M}_{\rm env}$ and $L_{\rm acc}$, we (initially) assume $\dot{M}_{\rm disk} = \dot{M}_{\rm env}$, and we adopt stellar radii, luminosities, and masses from the \citet{sie00} online models of pre--main-sequence stars at an age of $3\times10^5~{\rm yr}$.  The final conclusions are not sensitive to the exact stellar parameters chosen. 

HOPS 166 is modeled as a luminous star-disk system with a low-density envelope seen through a few magnitudes of visual extinction. Ê(The modeled inclination of $40^\circ$ is considered a lower limit; \citealt{eis94} find that the outflow associated with this source is close to the plane of the sky.)  The best-fitting central star from the Siess et al.\ models has a mass of $2.2~M_{\sun}$, implying an envelope infall rate of $4\times10^{-7}~M_{\sun}~{\rm yr}^{-1}$ and an accretion luminosity that is 20\% of the total luminosity. The low accretion luminosity and low envelope mass (inferred from the model parameters) of only $0.02~M_{\sun}$ imply that HOPS 166 is in the late stages of protostellar evolution and that the central star has accreted most of its mass.  (\citealt{chi01} classified this source as a deeply embedded Class 0 object based on SCUBA and IRAM mapping at 450, 850, and 1300 $\mu$m.)

In contrast, HOPS 168 is much more embedded and luminous than HOPS 166.  Its envelope mass is 2.7~$M_{\sun}$.  The implied stellar mass is $0.3~M_{\sun}$, the envelope infall rate is $3\times10^{-5}~M_{\sun}~{\rm yr}^{-1}$, and the accretion luminosity is more than 95\% of the total.  A star more massive than $0.3~M_{\sun}$ is possible if $\dot{M}_{\rm disk}<\dot{M}_{\rm env}$, meaning infalling matter is piling up on the disk, leading to episodic accretion \citep[e.g.,][]{vor05}.  For example, the central star could have a mass as high as $1.8~M_{\sun}$ if $\dot{M}_{\rm disk}=0.1~\dot{M}_{\rm env}$.

The two remaining protostars, HOPS 165 and HOPS 203, are separated by only 13\arcsec, or a projected 5500~AU.  The SED of HOPS 165 drops off precipitously beyond 30 $\mu$m. Ê This requires a very small, tenuous envelope and a low-mass disk. ÊThe flux from the moderately luminous star-disk system is seen behind 35 magnitudes of visual extinction.  Our interpretation is that HOPS 165 must be seen through the dense envelope of the nearby HOPS 203 ($M_{\rm env}=2.4~M_{\sun}$). ÊHOPS 203 itself is a 3\arcsec\ binary \citep{rod90}.Ê If the proximity of HOPS 165 is not due to chance, this region is home to a hierarchical multiple system of three protostars within a projected radius of 5500 AU. Accordingly, the small envelope size of the HOPS 165 model may result from its proximity to HOPS 203. ÊThe implied stellar mass of HOPS 165 is $1.4~M_{\sun}$, the envelope infall rate is $2\times10^{-7}~M_{\sun}~{\rm yr}^{-1}$, and the accretion luminosity is 10\% of the total.  On the other hand, the implied stellar mass of HOPS 203 is $0.1~M_{\sun}$, the envelope infall rate is $2\times10^{-5}~M_{\sun}~{\rm yr}^{-1}$, and the accretion luminosity is more than 95\% of the total.  Again, the central star may have a higher mass if $\dot{M}_{\rm disk}<\dot{M}_{\rm env}$.  We assume that the accretion is dominated by one member of the 3\arcsec\ binary, but the results will not change significantly if both accrete equally.  The 160~$\mu$m PACS measurement for HOPS 203 exceeds the fit by a factor of 2; this may be due to cold envelope material in our aperture that is not accounted for in our models.

We conclude that two of the protostars (HOPS 168 and 203) are in an active state of mass infall and accretion, while the other two (HOPS 165 and 166) have only residual envelopes.  This finding demonstrates {\it Herschel}'s unique and critical contribution to the audit of the flow of mass from the outer protostellar envelope onto the central protostar.

\begin{acknowledgements}
This work is based on observations made with the {\it Herschel Space Observatory}, a European Space Agency Cornerstone Mission with significant participation by NASA, and with the {\it Spitzer Space Telescope}, which is operated by the Jet Propulsion Laboratory, California Institute of Technology, under a contract with NASA.  Support for the {\it Herschel} and {\it Spitzer} analysis was provided by NASA through awards issued by JPL/Caltech.  We are grateful to Barbara Whitney and her collaborators for making their radiative transfer code available to the community.
\end{acknowledgements}

\end{document}